\documentstyle[prl,aps,epsf,floats]{revtex} 

\begin{document} 

\wideabs{ 

\title{Surprises in the doping dependence of the Fermi surface in (Bi,Pb)$_2$Sr$_2$CaCu$_2$O$_{8+\delta}$}
\author{A. A. Kordyuk$^{1,2}$, S. V. Borisenko$^1$, M. S. Golden$^1$, S. Legner$^1$, K. A. Nenkov$^1$, M. Knupfer$^1$, J. Fink$^1$, \\ H. Berger$^3$, L. Forr\'{o}$^4$}
\address{$^1$ Institute for Solid State and Materials Research Dresden, P.O.Box 270016, D-01171 Dresden, Germany}
\address{$^2$ Institute of Metal Physics of National Academy of Sciencies of Ukraine, 03142 Kyiv, Ukraine}
\address{$^3$ Institut de Physique Appliqu\'ee, Ecole Politechnique F\'ederale de Lausanne, CH-1015 Lausanne, Switzerland}
\address{$^4$ DP/IGA, Ecole Politechnique F\'ederale de Lausanne, CH-1015 Lausanne, Switzerland} 

\date{\today} 

\maketitle 

\begin{abstract} A detailed and systematic ARPES investigation of the doping-dependence of the normal state Fermi surface (FS) of modulation-free (Pb,Bi)-2212 is presented. The FS does not change in topology away from hole-like at any stage. The data reveal, in addition, a number of surprises. Firstly the FS area does not follow the usual curve describing $T_c$ vs $x$ for the hole doped cuprates, but is down-shifted in doping by ca. 0.05 holes per Cu site, indicating either the break-down of Luttinger's theorem or the consequences of a significant bi-layer splitting of the FS. Secondly, the strong $k$-dependence of the FS width is shown to be doping {\it independent}. Finally, the relative strength of the shadow FS has a doping dependence mirroring that of $T_c$.
\end{abstract} 

 }

The shape and topology of the Fermi surface (FS) of the high temperature superconductors (HTSC), and in particular of the Bi$_2$Sr$_2$CaCu$_2$O$_{8+\delta}$ (Bi-2212)-based systems, has been a hot topic from the very beginning of the HTSC era \cite{Dessau,Ma}, and is still the subject of lively discussion today \cite{Chuang,Fretwell,BorisPRL}. In the past, the existence of a large, hole-like FS in angle-resolved photoemission spectroscopy (ARPES) was taken as support for the validity of Luttinger's theorem for the superconducting cuprates \cite{Olson,Campuzano}. While some ARPES studies of Bi-2212 conclude that a large, hole-like FS persists even to very low doping levels \cite{Ding}, other data imply a change in FS topology \cite{SchwallerEPJ} or the presence of hole- pockets \cite{Marshall} at underdoping. Recent data from La$_{2-x}$Sr$_x$CuO$_4$ (LSCO) have been interpreted in terms of a change of FS topology from hole-like for $x<0.2$ to electron-like for higher doping levels \cite{Ino}.

The recent improvement in the performance of photoemission instrumentation (in particular in the angular resolution) has led to a renaissance in the direct determination of the basal plane projection of the FS using ARPES. Considering the fundamental importance of the FS topology and shape in deciding the physical properties of a solid, it is natural to want to study its doping dependence in the Bi-based HTSC directly and with high precision using high resolution FS mapping.

The ARPES experiments reported here were performed using monochromated He I radiation and an SES200 electron analyzer combined with a precise 3 axis sample rotation system. The overall resolution in ${\bf k}\omega$-space was set to 0.014 \AA$^{-1} \times$ 0.035 \AA$^{-1} \times$ 19 meV which are the FWHM momenta (parallel and perpendicular to the analyzer entrance slit) and energy resolutions, respectively \cite{BorisPRB}. The samples were cleaved in-situ to give mirror-like surfaces and all data were measured above the pseudogapped regime at 300K within 3-4 hours of cleavage. We investigated a set of high quality single crystals of Pb- doped Bi-2212 which had undergone different oxygen loading procedures. As we have pointed out earlier \cite{BorisPRL,BorisPRB,SibyllePRB}, it is wise to use the Pb-substituted variants for such experiments as these systems do not possess the incommensurate modulation of the BiO layers which in pristine Bi-2212 leads to the appearance of strong diffraction replicas of the main and shadow FS features in the maps, thus disqualifying a detailed discussion of the FS topology, shape and area as a function of doping. In the following, we label the samples, which span a $T_c$ range of 35 K around optimal doping according to their $T_c$: UD 76K, UD 85K, UD 89K, OD 81K, OD 72K and OD 69K (UD and OD stand for underdoped and overdoped).

\begin{figure*}[!t]
\begin{center}
\epsfxsize=18cm
\epsfbox{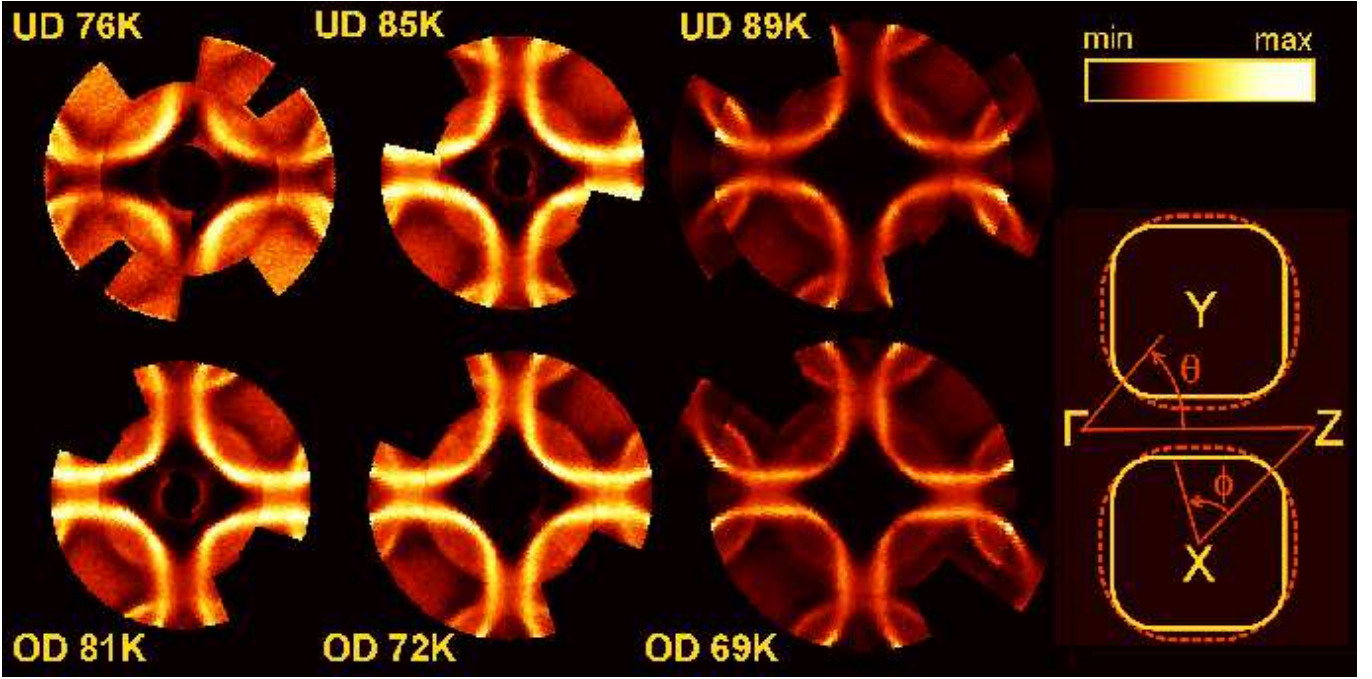}
\end{center}
\caption{Basal plane projection of the normal state (300K) Fermi surface of Bi(Pb)-2212 from high resolution ARPES. The $E_F$ intensity (normalized to the signal at $\omega$ = 0.3 eV) is shown in color. The $T_c$ of each sample is indicated. The sketch shows the FS for the OD 69K dataset defined by joining the maxima of fits to the normalized $E_F$ MDC's (yellow line).}
\label{maps}
\end{figure*}

Fig.\ref{maps} shows the Fermi surface maps for all six doping levels. Each dataset contains ca. 5000 ARPES spectra and the measured maps cover half of the area of each image shown in Fig.1. In this way we collect data from a significantly larger region of $\bf k$-space than the irreducible octant, which brings the advantage of enabling a quantitative correction of angular misalignments of the crystal to a precision of 0.1$^\circ$.

To minimize the effects of the factors separating the ARPES intensity distribution from the spectral function, the data were 'self-normalized' by dividing the signal from the Fermi level, $I(\bf k,\omega=0)$, by the signal at highest binding energy, $I({\bf k},\omega_{hbe})$ (here $\omega_{hbe}$ is 300 meV). The FS topology and shape derived from these data do not depend sensitively upon the use of any reasonable self-normalization denominators \cite{BorisPRB}.

Before going on to discuss the data in a more quantitative manner, we first cover what can be learned directly from a simple visual inspection of Fig.1. (a) There is no topological change of the main FS within the doping range studied - it remains hole-like (centered at the X,Y points), in contrast to recent data from the LSCO system \cite{Ino}. (b) As hole doping is increased, the main FS 'barrels' increase in size (as can easily be seen in the decrease of the inter-barrel separation around the M ($\pi$,0) point), accompanied by an increase in the size of the lenses formed by main FS and shadow FS (SFS). (c) The shape of the FS barrels changes from being quite rounded at low doping to take on the form of a square with well-rounded corners at higher doping. (d) The SFS exists at all doping levels.

We stress that these statements describe experimental observations and are independent of any particular data analysis or physical interpretation.

One of the fundamental questions in the physics of 2D strongly correlated electron systems is to what extent the interacting electron system can be described by models derived perturbatively from the non-interacting case. One way to test this is to consider the validity or otherwise of Luttinger's theorem, which can be paraphrased by stating that the volume (area in 2D) of the FS should be conserved upon switching on the interactions. Thus if we are able to pin down the doping dependence of the exact path in ${\bf k}$-space which represents the Fermi surface in, for example, the (Pb,Bi)-2212 HTSC without knowing, a priori, its shape, we would be able to evaluate the doping dependence of the FS area and thus test Luttinger's theorem. The best approach here is to locate the maxima in the $E_F$ momentum distribution curves (MDC's) describing tracks crossing the FS (preferably at right angles) \cite{BorisPRB,Aebi}.

Such a fitting procedure was carried out for the OD 69K sample. The detailed result is well described by a FS having the form of a square with rounded corners, which confirms the visual impression from the intensity map for this sample. A sketch of the fit result is shown as the yellow line on the right hand side of Fig.\ref{maps}. The FS maps from the other samples were then fitted, whereby the extent of the straight sections, as well as the size of the barrel as a whole were varied to optimize the fit to the data. We can then derive the hole concentration $x$ from the simple relation $x = S_b / \Gamma$X$^2 - 1$, where $S_b$ is the area of main FS barrel.

\begin{figure}[!t]
\begin{center}
\epsfxsize=8.47cm
\epsfbox{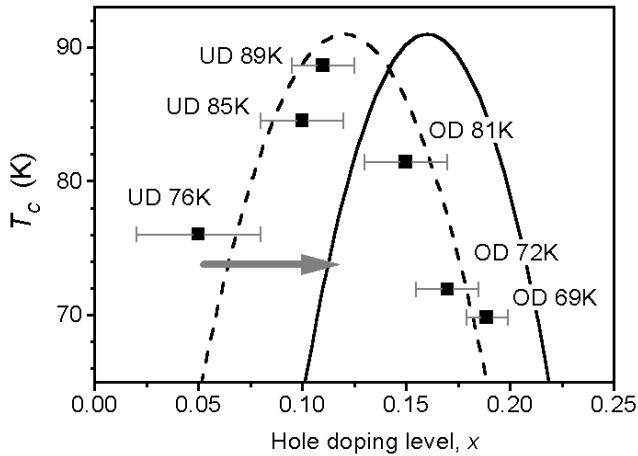}
\end{center}
\caption{Symbols: critical temperatures vs the hole concentration, $x_{FS}$, the latter being calculated directly from the area of the FS's shown in Fig.1. The solid line shows the commonly-used empirical relation for $T_c$ vs $x$ (Ref. [15]).}
\label{doping}
\end{figure}

The results obtained from the analysis of the FS area are shown in Fig.\ref{doping} in the form of a $T_c$ vs $x$ plot. The solid line shows the commonly employed empirical relation between $T_c$ and $x$ \cite{Tallon}. The surprising result here - for the six samples spanning a total of 35K in $T_c$ - is that the co-ordinate pairs matching the $T_c$'s to the doping level taken directly from the experimentally determined Fermi surface area also give a parabolic curve (shown as a dotted line), but that this curve is down-shifted in doping by ca. 0.05 towards the underdoped side of the phase diagram.

To explain the discrepancy, the following possibilities spring to mind: (i) the universal $T_c$ vs $x$ curve \cite{Tallon} is not valid for Pb-doped Bi-2212; (ii) the doping level at the surface is not the same as in the bulk; (iii) we are 'missing' a small amount of FS area in our analysis of the data; (iv) Luttinger's theorem is not valid for the doping range studied here.

The Bi-2212 family of HTSC offers few easy handles on the true hole doping level in the CuO$_2$ planes, as traditional sources of data such as Hall effect data are notoriously complicated in the HTSC and there is no controlling dopant (such as Sr in LSCO) nor simple-to-characterize oxygen reservoir (such as in YBa$_2$Cu$_3$O$_{7-\delta}$). Thus, it is quite possible that the dotted ARPES-derived $T_c$ vs $x$ parabola represents the true situation. However, the universality of the optimal doping level for many HTSC systems would argue against this explanation.

If the doping level at the surface were lower than in the bulk (for example by loss of oxygen at the surface), such a deviation should be strongly dependent on the oxygen loading procedure, affecting the OD samples more strongly than the UD, which is clearly not the case. Furthermore, the fact that the superconducting gap seen in ARPES data from the same samples (not shown), closes unambiguously at the {\it bulk} $T_c$ in the overdoped systems, is incompatible with a lower doping level at the surface.

By 'missing FS area' we indicate the possibility that the FS has a complex structure which is difficult to observe in ARPES experiments, such as the bi-layer splitting or BiO-derived pockets predicted in band structure calculations \cite{Freeman}. It has been shown recently that it is possible to resolve the bilayer splitting in ARPES studies of overdoped Bi-2212 \cite{Feng,Chuang2}. In this case, the blurring of the FS (see Fig.\ref{maps}) on going from the nodal to the antinodal point for all doping levels, which is often attributed to the complex physics of antinodal electrons (an absence of well-defined quasiparticles), could at least partially be due to the complex structure of the FS. In order to examine this possibility, in the following we analyze the FS width in more detail.

\begin{figure}[!t]
\begin{center}
\epsfxsize=8.47cm
\epsfbox{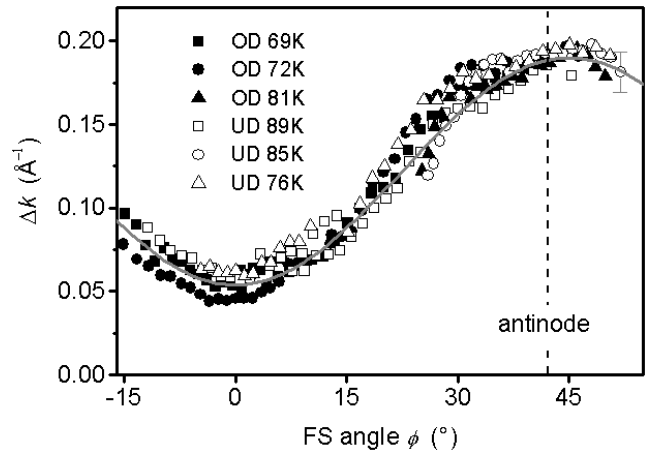}
\end{center}
\caption{The width of the main FS $\Delta k$ vs the FS angle $\phi$ defined with respect to the nodal line. The $T_c$'s are indicated and the solid grey line represents the relation $\Delta k (\phi) = \Delta k_0 + \Delta k_1 \sin^2(2\phi)$, for details see text.}
\label{width}
\end{figure}

In Fig.\ref{width}, we show the width of the FS, $\Delta k$ vs $\phi$, the latter being the angle away from the nodal line, as indicated in Fig.\ref{maps}. The $\Delta k$ values were derived from fitting $E_F$ MDC's using a Lorentzian profile with $\Delta \theta$ FWHM in angle and $\Delta k = |{\bf k}| \Delta \theta$. For all doping levels investigated the FS width is strongly $k$-dependent, being maximal near the antinode and minimal at the node. The dotted line in Fig.\ref{width} shows that the data can be well described by the function $\Delta k_0 + \Delta k_1 \sin^2(2\phi)$, where $\Delta k_0$ = 0.054 \AA$^{-1}$ and $\Delta k_1$ = 0.136 \AA$^{-1}$. Remarkably, the observed $k$-dependence of the FS width is essentially {\it independent} of the doping level. This is difficult to reconcile with a FS width determined solely by the complex physics of the FS electrons, as within such a picture the difference in the coupling to interactions between the nodal and antinodal regions should decrease continually as the doping increases.

On the other hand, exploiting the 'complex FS structure' scenario, we can assume the existence of a bi-layer splitting, $\delta k (\phi)$ at $E_F$. For the case in which the maxima of the MDCs (i.e. the intensity in a self-normalized FS map such as those of Fig.1) corresponds to the inner bi-layer split FS barrel, this would result in a shift of the observed doping level of $\delta x \approx \delta S / 2 \Gamma$X$^2$, where $\delta S \approx \left<k_b\right> \int_{0}^{2\pi} \delta k (\phi) d\phi$ is the difference in area between the split barrels, $k_b$ is the radius of main FS barrel with respect to the X-point ($\left<k_b\right> \approx 0.6 \Gamma$X) and $\Gamma$X = 1.161 \AA$^{-1}$. This effect is illustrated schematically in the cartoon shown in Fig. 1 where the yellow (red) barrels represent the smaller (larger) FS's resulting from the bi-layer splitting. Taking a Lorentzian form for the $E_F$ MDC which cuts the FS, we expect a bi-layer splitting induced FS width given by $\Delta k \approx W + 3(\delta k)^2 / 2 W$ (where $W$ is the FWHM of the FS without splitting and $\delta k \le W / \sqrt{3}$ is assumed to hold). In such a manner we can estimate an upper limit for $\delta x$ = 0.07, which is illustrated in Fig.\ref{doping} by the broad grey arrow. This demonstrates that the effect of the bi-layer splitting is enough to explain the downshift of the $T_c$ vs doping parabola.

\begin{figure}[!t]
\begin{center}
\epsfxsize=8.47cm
\epsfbox{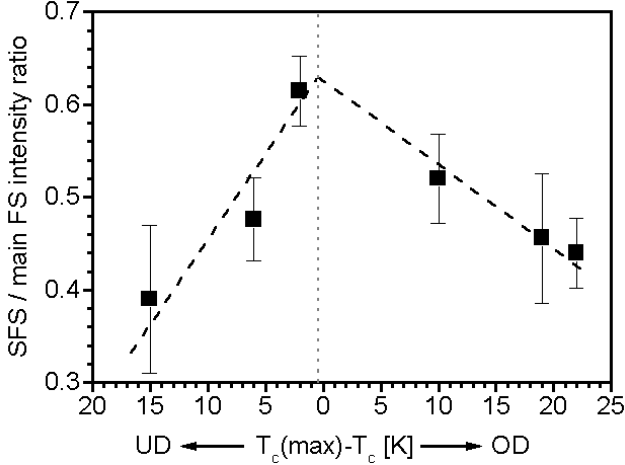}
\end{center}
\caption{SFS to main FS intensity ratios vs $T_c^{max} - T_c$. The dashed straight lines are guides to the eye.}
\label{shadow}
\end{figure}

Finally, we note in this context that the upper limits of $\delta k$ (and subsequently $\delta x$) obtained above correspond to the Rayleigh resolution limit at which $\delta k = W / \sqrt{3}$. This same limit also defines the lower bound for the $\phi$-dependence of the Fermi surface width which arises from sources other than the bi-layer splitting: $\Delta k (\phi) / \Delta k_0 = 1 + 1.3 \sin^2(2\phi)$. Thus, in considering either the 'complex physics' or 'complex FS structure' scenarios we discuss two extremes, whereas the real situation may well include contributions from both. For example, at high hole doping, the $\phi$-dependence of $\Delta k$ from 'complex physics' should flatten out, which would be counteracted by the increasing bi-layer splitting for this doping regime (in which the flat bands approach closer to $E_F$). Conversely, at low hole doping, the $\phi$-dependence of the coupling to interactions is strong, whereas the bi-layer splitting would be expected to be weaker. In this way we end up with the observed overall doping independence of $\Delta k(\phi)$.

As mentioned above, it is possible to compensate for the downshift of the $T_c$ vs. $x$ parabola in Fig.2 by taking the bi-layer splitting into account. It would then follow that the area of the main ARPES FS scales with (1+$x$) in holes across the complete doping range studied. This behaviour is in contrast to what is seen in transport measurements. Resistivity and Hall effect data indicate that the transport characteristics scale with $x$ \cite{Batlogg,Ong}, even into the overdoped regime \cite{Alloul}. Although it is conceivable that only those mobile electrons which have relatively low coupling to other degrees of freedom contribute to the transport, it is surely more than coincidental that this proportion should be exactly $x$/(1+$x$). This fundamental difference between the transport data and the ARPES FS is a key question which deserves detailed theoretical attention.

Lastly, if none of the scenarios just described are appropriate, then we must conclude that Luttinger's theorem is violated in these systems. We note that calculations of the 2D $t$--$J$ model \cite{Putikka} for 20\% hole doping have indicated a violation of Luttinger's theorem.

A final surprise that the FS has in store for us is shown in Fig.\ref{shadow}, in which the doping dependence of the intensity ratio of the SFS to that of the main FS is plotted. The intensities were taken in each case from the same azimuthal MDC scan: i.e. with the same $|{\bf k}|$ value, some 0.13 \AA$^{- 1}$ from the point at which the SFS and main FS 'cross'. As Fig.\ref{shadow} shows, the SFS/FS ratio decreases not only on going from optimal to overdoping, but also on going towards the underdoped side of the phase diagram (the rate of change is, in fact, even faster on the UD side). This is in contrast to predictions based on an antiferromagnetic origin of the SFS \cite{Haas}, but could rather signal that the microscopic origin of the SFS is related to high $T_c$ superconductivity itself. Further work is needed, both on the experimental but also on the theoretical side, before the question of the shadow Fermi surface can be considered as solved.

In conclusion, we have presented a detailed and systematic ARPES investigation of the doping dependence of the normal state FS of the Bi-2212 family of HTSC materials. The data clearly show no change in the FS topology away from hole-like at any stage (from UD 76K to OD 69K). An analysis of the main FS area gives a parabolic $T_c$ vs $x_{FS}$ relation, shifted to lower $x$ by some 0.05 compared to the 'universal' relation \cite{Tallon}, which can be accounted for by the presence of two (unresolved) FS's near ($\pi$,0) due to a bi-layer splitting with a maximum value ca. 0.05 \AA$^{-1}$. Were the bi-layer splitting not responsible for this discrepancy, then Luttinger's theorem would appear to be violated in these systems. Furthermore, the FS width is shown to be strongly dependent on $k$, but for each particular $k_F$ point it is essentially independent of the doping level, which can be understood as a combination of the effects of the bi-layer splitting (dominating at higher doping) and the complex physics of the FS electrons (dominating at lower doping). Finally, the shadow FS is clearly visible for all doping levels, and has maximal intensity at optimal doping, raising the question of a possible link between the origons of the shadow FS and of superconductivity.

We are grateful to the the BMBF (05 SB8BDA 6) and to the Fonds National Suisse de la Recherche Scientifique for support and to S.-L. Drechsler, A. N. Yaresko, A. Ya. Perlov, R. Hayn, N. M. Plakida, and M. Eschrig for stimulating discussions.

\end{document}